\documentclass[ acmtog,
                authorversion   =true, 
                screen          =true,
                authordraft     =false, 
                review          =false, 
                timestamp       =true, 
                anonymous       =false]{acmart}
\AtBeginDocument{%
  \providecommand\BibTeX{{%
    \normalfont B\kern-0.5em{\scshape i\kern-0.25em b}\kern-0.8em\TeX}}}

\setcopyright{acmcopyright}\acmJournal{TOG}
\acmYear{2021}\acmVolume{40}\acmNumber{6}\acmArticle{271}\acmMonth{12} \acmDOI{10.1145/3478513.3480516}

\usepackage{enumitem}
\usepackage[nice]{nicefrac}
\usepackage{amsmath}
\usepackage{multirow}
\usepackage{comment}
\usepackage{wrapfig}
\newcommand{\breakpage}{}
 
\newcommand{\ttrans}{{T}}
\newcommand{\nc}{\vartheta} 
\newcommand{\tnc}{t_\nc}
\newcommand{\pnc}{p_\nc}
\newcommand{\pncc}{{p_{\Theta}}}
\newcommand{\msub}[1]{\scriptscriptstyle{\text{#1}}}
\newcommand{\mb}[1]{{\boldsymbol{#1}}}
\newcommand{\bb}[1]{{\mathbb{#1}}}
\newcommand{\pare}[1]{{\left(#1\right)}}

\newcommand{\rem}[1]{}

\newcommand{\norm}[1]{\left\lVert#1\right\rVert}
\newcommand{\inner}[2]{{\langle #1, #2 \rangle}}

\newcommand*\dd{\mathop{}\!\mathrm{d}}
\newcommand*\st{\mathop{}\!\operatorname{s.t.}}

\newcommand{\etal}{\textit{et al.~}}

\newcommand{\matlab}{\textsc{Matlab}}
\newcommand{\tod}[1]{\bar{#1}}

\newcommand{\revision}[1]{{\color{black}#1}} 

\graphicspath{{figures_lowres/}}

\newcommand\pmat[1]{\begin{pmatrix}#1\end{pmatrix}} 
\newcommand\bmat[1]{\begin{bmatrix}#1\end{bmatrix}} 
 
\begin{document}

\title{Generalized Deployable Elastic Geodesic Grids}

\author{Stefan Pillwein}
\orcid{0000-0003-4045-3867}
\affiliation{
  \institution{TU Wien}
  \country{Austria}
}
\email{stefan.pillwein@tuwien.ac.at}
\author{Przemyslaw Musialski}
\orcid{0001-6429-8190}
\affiliation{
    \institution{NJIT}
    \country{USA}
}
\email{przem@njit.edu}

\renewcommand{\shortauthors}{Stefan Pillwein and Przemyslaw Musialski}

\begin{abstract}
Given a designer created free-form surface in 3d space, our method computes a grid composed of elastic elements which are completely planar and straight. Only by fixing the ends of the planar elements to appropriate locations, the 2d grid bends and approximates the given 3d surface. 
Our method is based purely on the notions from differential geometry of curves and surfaces and avoids any physical simulations. 
In particular, we introduce a well-defined elastic grid energy functional that allows identifying networks of curves that minimize the bending energy and at the same time nestle to the provided input surface well. 
Further, we generalize the concept of such grids to cases where the surface boundary does not need to be convex, which allows for the creation of sophisticated and visually pleasing shapes. 
The algorithm finally ensures that the 2d grid is perfectly planar, making the resulting gridshells inexpensive, easy to fabricate, transport, assemble, and finally also to deploy. Additionally, since the whole structure is pre-strained, it also comes with load-bearing capabilities. 
We evaluate our method using physical simulation and we also provide a full fabrication pipeline for desktop-size models and present multiple examples of surfaces with elliptic and hyperbolic curvature regions. 
Our method is meant as a tool for quick prototyping for designers, architects, and engineers since it is very fast and results can be obtained in a matter of seconds. 
\vspace{40pt}
\end{abstract}




\begin{teaserfigure}
\includegraphics[width=\textwidth]{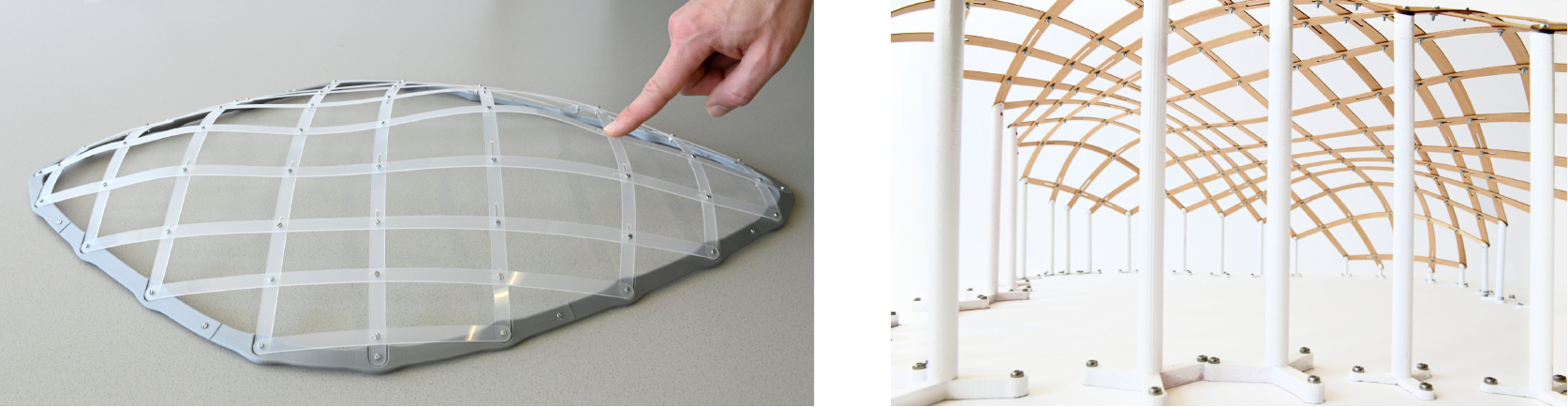}
\caption{Results of the generalized elastic grid method. Left: an approximation of a doubly-curved surface with elliptic and hyperbolic regions using an elastic geodesic grid, deployed from a perfectly planar state by fixing its boundary elements to an anchor-ring. Right: a closeup photography of an elastic geodesic grid dome which is also doubly-curved, fixed to 3d printed anchors. Please refer to Figure \ref{fig:results} for more examples. }
\label{fig:teaser}
\end{teaserfigure}

\maketitle

\vspace{10pt}
\section{Introduction}\label{sec:introduction}

The pursuit of novel structures suitable for design and engineering has been a very old topic in science and engineering. One of the ultimate goals of this race are structures that are light, cheap, and strong.

An interesting class of structures that largely fulfill these requirements surprisingly seems to have contradictory properties: 
Under load, they leave their stable state and compensate it with buckling. 
In mechanical terms, they elastically move to the post-buckling regime.
Usually, engineers want to avoid any buckling effects as they are regarded as structural failures. On the other hand, by adequately designing the structure's geometry, this seeming disadvantage can be of great practical use with a high level of efficiency. Especially, slender structures like rods, plates, or shells are suitable for design with such elastic bending techniques. 

In practice, the post-buckling effect can be observed on a simple igloo tent whose shape is maintained by two poles bent over each other. This construction has many advantages; it is compact, lightweight, and easy to build, but still effective and durable. 

In the past, the possibilities for the design of free-form structures were very limited, and the material has almost always imposed the final shape. The design of sophisticated geometric shapes was too complex without advanced computational tools. Nonetheless, the paradigm fascinates and has been applied in many scales and domains: from large-scale architecture~\cite{Shukhov1896}, over medium-scale furniture design~\cite{Panagoulia2016}, to mesoscale structures in material science~\cite{Lavine2015}. 

Recently, several approaches for the computational design of deployable elastic structures~\cite{Soriano2019,Panetta2019,Pillwein2020} have been proposed.
Nonetheless, while these methods come with excellent results, they suffer from a number of limitations. For instance, Panetta~\etal~\shortcite{Panetta2019} introduce elastic gridshells composed of elements with varying cross-sections and allow for varying boundaries, however, their approach does not always ensure the planarity of the 2d configurations. 

In contrast, the method of Pillwein~\etal~\shortcite{Pillwein2020} generates perfectly planar 2d layouts, on the downside, their method is limited to scissor-like convex quadrilateral patches only and does not allow free-form boundaries. While more complex shapes are possible by stitching multiple quadrilateral patches together~\cite{Pillwein2020a,pillwein2021}, an arbitrary boundary is still not possible. 

In this paper, we introduce a new method to deal with arbitrary, even non-convex boundaries.
Our method is based on assumptions derived from differential geometry of curves and surfaces, similar to the model of Pillwein~\etal~\shortcite{Pillwein2020}, where the elastic grid follows geodesic curves on the surface and expects that the lattice members are bendable only along a single axis. 
While this poses a certain limitation on the surfaces that can be approximated, it still allows exploring a rich space of possible designs with double curvature, especially if the surfaces have non-convex boundaries. 

Our model allows us to compute the grids using purely geometric notions with no need for physical simulation, making it computationally very efficient. At the same time, the proposed geometric concepts are well-founded by the theory of minimal energy curves and we can show a very close match of our results to the outcomes of physical simulation performed with the state-of-the-art discrete elastic rods model \cite{Bergou2008}. 

Indeed, most of our concepts can be reduced to the computation of geodesic distances on the surface, including intersections of curves, and not even the computationally expensive tracing of the paths of the curves is necessary. Additionally, we extend it fluently to the discrete domain, which allows for efficient updates of the combinatorics of the grids. 

\begin{figure*}
\centering
\includegraphics[width=\textwidth]{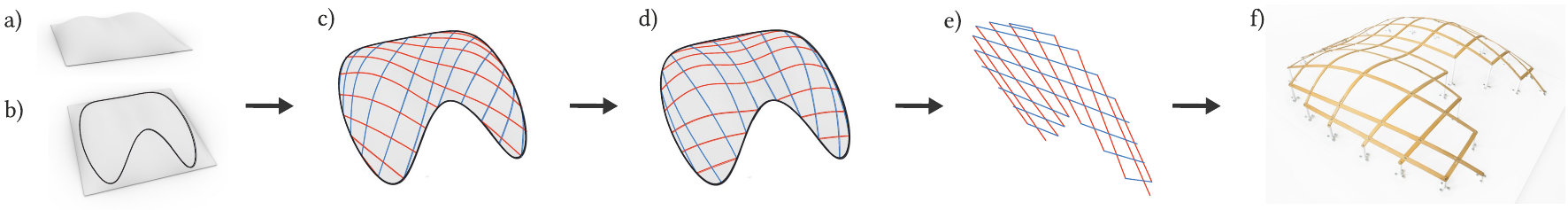}
\caption{Overview of our modeling pipeline: (a) initial surface, (b) designer chooses a boundary curve on the surface, (c) initial grid (designer provides an arbitrary number of members),  (d) grid optimization (we find a grid that nestles to the surface well), (e) planarized grid, (f) fabricated gridshell.  }
\label{fig:overview}
\end{figure*}

In particular, the contributions of the paper are the following: 
\vspace{-10pt}
\begin{itemize}

\item An elastic grid on the surface should approximate the surface well and capture its characteristics. The task of the choice of proper members is non-trivial and has been solved by Pillwein~\etal~\shortcite{Pillwein2020} using a geometrically driven heuristic. We introduce a well-defined energy functional which allows us to identify so-called \textit{least-effort} and \textit{most-effort} curves on the surface, which ensure to capture the surface characteristics well and provides aesthetic grid layouts. 

\item More sophisticated surfaces exhibit boundaries which are non-convex (c.f.  Fig.~\ref{fig:overview}), in fact, they can possess multiple non-convex regions. This implies that the planar configuration becomes more intricate, hosting sub-families of members,
and poses the challenge of varying connectivity grids. We propose an efficient algorithm based on distance computations only that updates the combinatorics of the grids. 

\item 
A crucial requirement of deployable structures is that the undeployed state remains perfectly planar. This requirement is important for the ease of fabrication, transportation, and assembly and should not be underestimated---dealing with bent elements is considerably more difficult and more expensive than with planar ones. Therefore, we introduce a generic planarization algorithm that also takes fabrication constraints into account. 

\item Finally, we introduce a digital fabrication pipeline for the grids and present a number of our results as desktop-size models fabricated from wood or acrylic glass as a proof of concept of our approach. 
\end{itemize}

In the remainder, our paper is organized as follows: In the next section, we review related works, in Section~\ref{sec:background} we discuss the goals and assumptions of our approach.
In Section~\ref{sec:representation} we describe the concepts of the representation of the grids, and in Section~\ref{sec:gridgeom} we discuss the computation of finding grid layouts with non-convex boundaries. 
In Section~\ref{sec:grid_energy} we describe the background of the elastic energy which we use for finding optimal grids. 
In Section~\ref{sec:grid_planarization} we propose an optimization algorithm for the planarization of spatial grids. 
Finally, in Section~\ref{sec:results} we present quantitative and qualitative results, including fabricated models, and in Section~\ref{sec:discussion} we discuss the limitations and conclude our work. 

\section{Related Work}

    \paragraph{Gridshell Structures}
    Structures that gain their strength and stiffness through their curvature have been used in architecture and design since ancient times~\cite{Lienhard2013}. At the end of the 19th century, Shukhov applied the idea for the Rotunda of the Panrussian Exposition \cite{Shukhov1896}, and it was further pursued by famous architects, e.g., by Frei Otto for the construction of the roof of the Multihalle at the Mannheim Bundesgartenschau \cite{Happold1975}. 
    
    This form of structure erection has been summarized in the architecture and construction literature as the \textit{active bending} paradigm \cite{Lienhard2013,Lienhard2018}.  Modern and easy-to-use computational methods increased the interest of the scientific community in systematically utilizing elastic bending to realize curved shapes. Until recent advances in computer science, they could only be form-found empirically \cite{gengnagel2013active}. 
    Architectural works aim at the approximation of gridshells and combine lightweight structural design with aesthetics \cite{Soriano2015,Soriano2017}. 
    
    Existing design approaches are often based on particular kinds of surface curves, e.g., asymptotic curves \cite{Schling2018}. 
    Besides their aesthetic qualities, such structures can also be assembled in initially flat segments. They naturally transform to a curved state by their internal forces, however, such curve networks can only be found on surfaces with negative Gaussian curvature.

    \paragraph{Deployable Structures}
    Much research is currently being carried out on gridshell-structures that can be deployed. They can be classified based on their deployment mechanism: inscribed in a grid that is deployed \cite{Panetta2019,X_shells_pavilion,Soriano2019,Pillwein2020} or by other external mechanisms like inflatable air cushions \cite{pneumatic,Konakovic-Lukovic2018}.
    We are interested in the first case. The design approaches of inscribing the deployment mechanism into the grid, however, differ: 
        
    To create an X-Shell \cite{Panetta2019}, a planar grid layout is designed using curved or straight members and actuated via physical simulation. In multiple layout iterations which do not require a target surface, the designer finds a satisfactory shape by adapting the planar design. Target surfaces can, however, be approximated using shape optimization (presuming a good planar initialization). The practical feasibility of these structures was investigated with the construction of a pavilion \cite{X_shells_pavilion}.

    The G-Shells approach \cite{Soriano2019} proposes to planarize a specific geodesic grid using physical simulation and an evolutionary multi-objective solver. This induces a geometric error, so the flat grid cannot be deployed to match the geodesic grid perfectly but still create beautiful shapes. However, the space of realizable shapes is not fully characterized. 
    
    In contrast to the former approaches, elastic geodesic grids \cite{Pillwein2020} use the concept of notches, which prevents the geometric error in the grid layout and allows for a very close approximation of a target surface. However, it also makes deployment more complicated as a sliding of members is necessary. The method takes a design surface as input and produces a deployable grid layout without using physical simulation. 
    
    Besides elastically bendable structures, there has also been extensive research on various deployable structures and deployment mechanisms methods. 
    One way to easily construct spatial shapes from flat sheets is by appropriately folding paper \cite{Massarwi2007a,Dudte2016}, which is inherently related to the Japanese art of Origami. 
    Also Kirigami, a technique to cut patterns into planar sheets to allow solid faces to rotate about each other, deforming in three dimensions while remaining planar has been explored for deployable surfaces~\cite{Liu2020,Jiang2020} and recently also bi-stable structures~\cite{Chen2021}.

    \vspace{-5pt}
    \paragraph{Bendable and Stretchable Structures}
    
    Elastic deformation of surfaces based on variational principles of minimal energy has a long research history in the computer graphics community \cite{Terzopoulos1987,Welch1992}. 
    These approaches usually assume that the structure can elastically bend and stretch.

    For instance, programmable elastic structures are based on both bending and tensile energy, e.g., by using prestressed latex membranes to actuate planar structures into free-form shapes \cite{Guseinov2017}. This method has been extended to programmable material sheets composed of mesostructures and membranes to design materials that stretch and bend to evolve to doubly-curved surfaces over time \cite{Guseinov2020}. 
    Another approach is to combine elastic rods and membranes leading to Kirchhoff-Plateau surfaces that allow easy planar fabrication and deployment \cite{Perez2017a}. 
    Furthermore, flexible rod networks \cite{Perez2015}, which additionally allow for controllable elastic deformation of given shapes, have been explored.
    
    A combination of precomputed flexible meso-cells leads to a method where planar configurations can deform to desired shapes if appropriate boundary conditions are applied \cite{Malomo2018a}.
    This approach has been tested by constructing a pavilion \cite{laccone2019flexmaps} on an architecturally relevant scale. 
    
    Another technique, called tensegrity, is to combine elastic and stiff elements to create physically stable structures, which has been recently explored for computational design~\cite{Pietroni2017}.  
    Also recently, a method for the design of kinetic wire characters has been proposed~\cite{Xu2018}, where custom springs are introduced to adapt the stiffness of the wires. 

    In fact, much attention has been paid to the design of doubly-curved surfaces, which can be deployed from planar configurations due to the ease of fabrication. One way of achieving this goal is by using auxetic materials \cite{Konakovic2016} which can nestle to doubly-curved spatial objects, or in combination with appropriate actuation techniques, can be used to construct complex spatial objects \cite{Konakovic-Lukovic2018}. 
    
    \revision{
    Surface-based inflatable structures \cite{panetta2021} utilize expanding tunnels fabricated by fusing two layers of thin material to approximate surfaces. Layouts are found by including the bending energy stored in the tunnels using the shape operator, a method that is closely related to how we express the energy of lamellae.
    }

    \paragraph{Bendable Non-Stretchable Structures}
    
    In contrast to methods that allow bending and stretching, our approach assumes that the elastic elements can bend and twist but not stretch and must therefore maintain the same length in the planar and spatial configuration. 

    Mappings of geodesic nets on a surface onto geodesic nets on a different surface (including the plane) were a topic of classic differential geometry \cite{Voss, Lagally}. It has been shown that arc-length preserving mappings of continuous geodesic nets onto each other require rhombic geodesic nets, i.e., need a parameterization of the surface with the net curves as parameter curves and \begin{math} E = G \end{math} in the fundamental form. The resulting  Liouville surfaces are very limited in shapes, and therefore not useful for our free-form design purpose. 
    
    A lot of attention has been paid to the approximation of free-form surfaces using developable surfaces \cite{Pottmann2010} which can be fabricated from 2d flat elements by cutting. 
    By bending and combining 2d elements, complex free-form surfaces can be erected. 
    On the theoretical side, a novel representation of developable surfaces using quadrilateral meshes with appropriate angle constraints~\cite{Rabinovich2018} or a definition of developability for triangle meshes~\cite{Stein2018} have been proposed recently. Also, optimal piecewise wrapping of doubly-curved surfaces with developable patches has been explored~\cite{Ion2020a}. 
    
    Other works deal with curved folds and efficient actuation of spatial objects from flat sheets \cite{Kilian2008,Kilian2017}. Discrete geodesic parallel coordinates have been introduced for modeling of developable surfaces~\cite{Wang2019}. 
    Another related work combines the idea of elastic bending and weaving. This allows for creating physically governed surfaces woven out of foliations whose leaves approximate geodesic curves~\cite{Vekhter2019}.
    
    Also related to our work are approximations of surfaces based on Chebyshev nets \cite{Garg2014}, which have been further analyzed for their elastic characteristics~\cite{Baek2018}. Recently this method has also been used for the simulation of hemispherical elastic gridshells \cite{Baek2019}. 
    
    \revision{Introducing elements with variable stiffness furthermore enables a wide range of target equilibrium shapes. This space was recently rigorously characterized \cite{hafner2021} for elements that are only constrained at their boundaries, using a method to determine physically viable shapes by visual inspection. To find viable grid layouts, we face a similar task and build upon these insights.}  
    

\section{Goals and Assumptions}\label{sec:background}

\subsection{Objectives}\label{sec:objectives}

Our goal is to find geodesic grids on given free-form surfaces such that they can be realized as physical grids composed of slender physical elements, such as rods or strips, where the ratio of the cross-sections of the elements is about 1:10 with a distinct weak axis. In our experimental models, we use thin lamellae laser-cut from wooden panels or acrylic glass plates.

The grids can be deployed from 2d planar states to surfaces in 3d space by compressing and fixing the outer boundary of the planar state. The intuition is that hence the geodesic distance between the ends of the incompressible elements is longer than the distance between the ends in the embedding space, they will undergo buckling and take a shape that minimizes the bending energy along their length. Due to the global interaction of the members, properly connected at their intersections, a spatial structure emerges. 

The input surfaces are allowed to have general boundary curves, which can even be non-convex (cf. Fig~\ref{fig:overview}), such that various configurations are possible. 
The grids are erected by fixing the free ends of the planar elements to given anchors distributed along a boundary curve. This provides the constraints necessary to adopt the desired minimal potential energy state.
Note that anchors provide both the positions (function values) and the tangent planes (first-order derivatives) at the boundary points of the grid, but the tangent directions are not fixed (refer to Figures ~\ref{fig:results} and \ref{fig:results_additional}).

\subsection{Assumptions and Simplifications}

\begin{figure}
\centering
\includegraphics[width=\columnwidth]{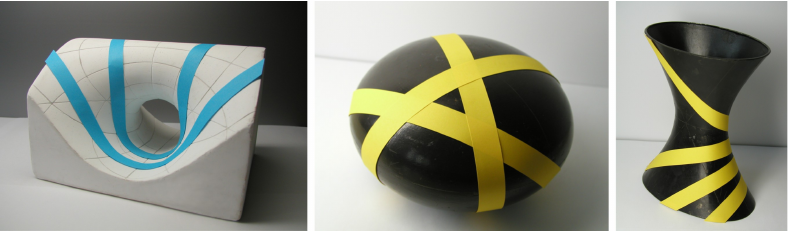}
\caption{Examples of paper strips ``glued'' to the surfaces. The centerlines of the strips follow geodesic curves on each surface. A sufficiently small strip can follow any geodesic on a surface but only a subset of all possible geodesics on a surface are suitable for elastic geodesic grids.  }
\label{fig:developable_strips}
\end{figure}

We minimize the error between the given target surface and the geodesic grid implicitly by assuming that the lamellae will behave very closely to geodesic curves on the target surface. 
This assumption is based on the fact that a geodesic curve has zero geodesic curvature $\kappa_g$ and exhibits only normal curvature $\kappa_n$, which is further discussed in Section~\ref{sec:grid_energy}. 

The intuition behind this assumption is that we can ``glue'' a thin strip along its centerline along such a curve on the surface and their lengths will match (cf. Figure~\ref{fig:developable_strips}).  
A sufficiently small strip can follow any geodesic on the surface, however, if the strip itself is elastic and resists bending, only a subset of all possible geodesics on the surface are suitable for elastic geodesic grids. Thus, our major goal in this paper is to find such grids to approximate the surfaces efficiently.

Another simplification is that since our computations are based purely on geometry, we do not take any physical quantities, like gravity or friction, into account. Our model also assumes perfectly geometrically non-linearly bendable materials, and we do not account for any material failure if the elastic region is left, resulting in severe structural failure. We do so since our model is meant for rapid form-finding and the generation of prototypes, however, such engineering constraints could be added easily on top of our model, if necessary.

\subsection{Representable Surfaces}\label{sec:surface_limitations}

The uniqueness of shortest geodesics is connected to the Gaussian curvature $K$ of the surface. Figuratively speaking, if a region of the surface has very high $K$, shortest geodesics will go around it and cease to be unique, which we need to prevent. Pillwein \etal \shortcite{Pillwein2020} investigated this problem and proposed an iterative smoothing procedure until the shortest geodesics between points on the boundary of a surface are unique. We use the same procedure, please refer to the paper for details.

\begin{figure*}
\centering
\includegraphics[width=\textwidth, trim=0 0 0 0, clip]{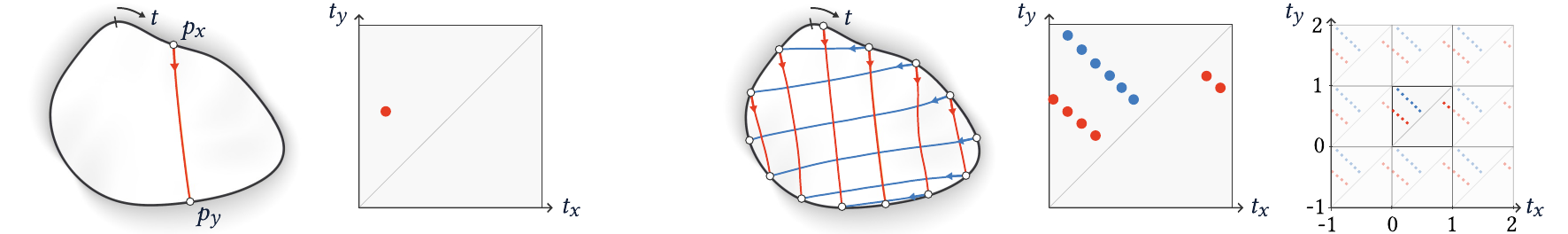}
\caption{
Our concept of grid representation in the dual space. We parameterize the boundary with constant speed, $t \in [0,1)$, and represent geodesics by the $t$-values of the endpoints.
Left: The geodesic between $p_x = p(t_x)$, $p_y = p(t_y)$ corresponds to a single point in the dual space, the direction $t_x \to t_y$ is also encoded.
Right: Two families of geodesics are represented in the dual space, spacing and directions of members define two sequences of points. If we extend the parameterization of the boundary to $(-\infty,\infty)$, the representation in the dual space is extended periodically.}
\label{fig:tspace}
\end{figure*}

\revision{
A stable grid on a surface with positive and negative $K$ requires geodesics between all regions of positive $K$ which are physically viable, i.e., the equilibrium shapes of the corresponding lamellae need to be close to the surface.
If the input surface has no inner bumps (extrema of $K$ far from the boundary), we can be pretty optimistic about finding a suitable grid.  }
If there are inner bumps, finding a grid that remains stable in the desired shape may be challenging. A sufficient but not necessary condition is based on the mean curvature $H>0$, which restricts the surfaces to the ones that can be achieved by inflating a balloon, i.e., excludes bumps that point inward~\cite{Konakovic-Lukovic2018}. 

\revision{
However, this condition is too strict for our purposes, as we discuss in Section~\ref{sec:grid_energy} and show in our results in Section~\ref{sec:results}, because it does not account for the interaction of grid members, which stabilize each other. }

\breakpage

\section{Grid Representation}\label{sec:representation}

This section describes the details of our implementation, grid representation, parameterization, and discretization. 
For the readers convenience, we first summarize the notation further used in the paper: 
\begin{itemize}
    \item $S\colon \bb{R}^2 \rightarrow \bb{R}^3$ with $(u,v) \mapsto S(u,v)$ is the input surface. 
    \item $p(t) \subset S$ is the closed boundary curve.
    \item $t$ is the unit speed curve parameter, w.l.o.g., $t \in [0,1)$. 
    \item $c(t_x,t_y) = c(p_x,p_y) \subset S$ represents the shortest curve connecting $p_x = p(t_x)$ and  $p_y = p(t_y)$.
    \item $d(p_x,p_y) = d(c(t_x,t_y)) \in \bb{R}_+$ denotes the distance between two points on the boundary.
    \item $g=\{c_1,\dots c_n\}$ and $h=\{c_1,\dots c_m\}$ are the families of curves of a grid; we order the elements by an increasing value of $t_x$. 
\end{itemize}

\subsection{Input and Output}\label{sec:method_overview}

The input to our system is a surface patch which is defined by a closed unit length boundary curve $p(t) \subset S$ on a surface $S \subset \bb{R}^3$. 
The curve is parameterized by the parameter $t$ w.l.o.g. in the interval $t \in [0,1)$. 
Both the surface $S$ and the curve $p$ are designer-provided. 

If the input surface is represented in parametric form (e.g., NURBS), it is tessellated to a polygonal mesh with sufficient resolution. Further on, our system works with triangular meshes. 
For several operations, we resort to a parameterization of the surface $S\colon \bb{R}^2 \to \bb{R}^3$, hence we expect the mesh to be well-paramete\-rized. We propose either utilizing the existing parameterization or employing other parameterization algorithms (e.g., least squares conformal mapping \cite{levy2002}). 

The output of our system is a planar grid composed of two families of lines denoted as $g$ and $h$ that cross each other. In the case of a non-convex boundary, any of the families can be further split into one or more groups, forming subfamilies; nevertheless, the grid pattern is always maintained, i.e., at each intersection exactly two members of either family cross each other. Please refer to Figures~\ref{fig:curve_families} and \ref{fig:member_correction} for a depiction.  

During the deployment, all free ends of the planar grid are fixed to the anchors, which provide them their location in space as well as their tangent planes. The tangent directions are enforced by the overall equilibrium state of the deployed grid. Note that additional constraints, like tangent directions on the free ends, could be prescribed to increase the structure's stability.

\subsection{Grid Parameterization}

\begin{figure}[b]
\centering
\includegraphics[width=\columnwidth]{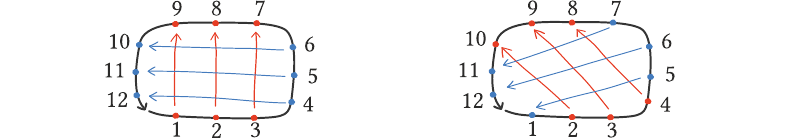}
\caption{Two families of grid members, their parameterization and order in the grid. By ordering the members in ascending order of their ``footpoints'', we can ensure a consistent pointing direction of members, even if they move along the boundary.  }
\label{fig:curve_families}
\end{figure}

We represent the geodesic grid by defining pairs of points $p_x = p(t_x)$, $p_y = p(t_y)$ on the boundary curve which connect to curves $c_i(t_x,t_y) = c_i(p_x$, $p_y)$. 
These curves represent the grid members and are organized in two families $g=\{c_1,\dots c_n\}$ and $h=\{c_1,\dots c_m\}$ such that members of one family can intersect the members of the other family as depicted in Figure~\ref{fig:curve_families}.
We order the elements of each family by an increasing value of $t_x$, which ensures that their pointing direction in a grid is consistent; in other words, it ensures that $p_x$ has a lower t-value than $p_y$, i.e., $t_x<t_y$. 

This parameterization allows us to express curves $c(t_x, t_y)$ which connect  $p_x$ and $p_y$ on the boundary as points in a 2d dual space. 
Since the boundary curve $p(t)$ is closed, the mapping is a symmetry group, such that for any function $f$, the following holds: 
\begin{align*}
f(t_x,t_y) = f(\ttrans(t_x,t_y))
\end{align*}
with
\[
\ttrans(x,y) = \bmat{0&1\\1&0}\bmat{x\\y}+\bmat{d\\d}, \quad \forall\, x,y\in \mathbb{R}\,, d\in \mathbb{Z} \,.
\]
Note that values $f(t_x,t_y)$ are translation and reflection-symmetric w.r.t. the affine transformation $\ttrans$, i.e., move along the borders in a periodic fashion and mirror across the diagonal. 
Points on the diagonal (i.e., $t_x=t_y$) 
represent infinitely short curves, i.e., points on the boundary.
We confine the dual space by  $[0,1)\times[0,1)$, with the symmetry given by $\ttrans$, please refer to Figure~\ref{fig:tspace} for a depiction.

\breakpage

\section{Grid Layout and Non-Convex Boundary}\label{sec:gridgeom}
In this section, we discuss the geometric approach to finding appropriate grid combinatorics, which is purely based on considering (geodesic) distances of points on surfaces. 

\subsection{Distance Fields and Distance Map} \label{sec:distance_map}

While there are many efficient algorithms for computing geodesic distances \cite{Crane2020}, tracing the path of a shortest geodesic is computationally much more expensive, requiring backtracing, variational shortening of curves, or other methods.

Hence, we utilize the concept of distance maps~\cite{Pillwein2020}, which can be computed and stored efficiently. 
Distances from all points $p_x$ on the boundary to all other points $p_y$ on the boundary establish the distance map 
\begin{equation*}
 D\colon (t_x,t_y) \rightarrow \bb{R}_+  \quad \text{with} \quad (t_x,t_y) \mapsto d(p_x,p_y)
\end{equation*}
where
\begin{equation*}
    d(p_x,p_y) = d(c(t_x,t_y)) = \int_{c} \norm{\mb{c}'(s)}  \dd s \in \bb{R}_+
\end{equation*}
denotes the geodesic distance (i.e., the arc length) between two points $p_x$ and $p_y$ on the boundary curve  w.r.t. the metric of the surface $S$. 
Since the boundary is closed, the distance map is defined over the dual space and is also subject to the same symmetry group.

In our implementation, the distance map is sampled on the boundary of the input surface mesh, and the resolution of the mesh gives its resolution. In practice, it is around $60\times60$ vertices.
Since the surface is not altered, distances are stored in a matrix $D$ and reused for many further computations, e.g., to detect non-convex parts of the boundary (Section \ref{sec:non-convex}). 
Any algorithm that determines geodesic distances accurately is suitable for our method, we use the algorithm proposed by Qin \etal \shortcite{qin16}.

\subsection{Intersections of Geodesic Curves} \label{sec:members_resolution}

\begin{figure}[b]
\centering
\includegraphics[width=\columnwidth, trim=0 0 0 0, clip]{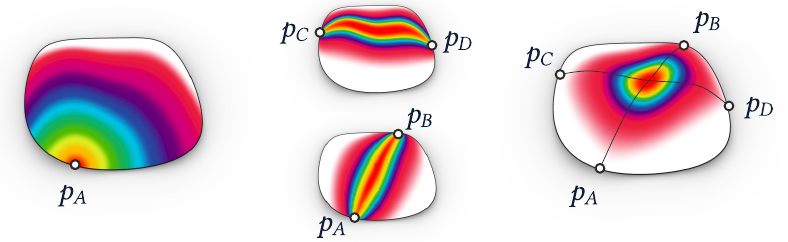}
\caption{We use distance fields to compute the intersection of two geodesics. Left: A distance field originating from $p_A$. Middle: When superimposing distance fields from $p_A,\,p_B$ and $p_C,\,p_D$, the paths of the geodesics can be recognized. Right: We recover the intersection of the geodesics by superimposing the distance fields and finding the minimum. }
\label{fig:distance_field}
\end{figure}

In our approach, geodesic paths are not explicitly computed, but we need to compute the intersections of members $\in g$ with other members $\in h$. We thus rephrase the problem of computing geodesic paths to computing intersections, using the link between shortest geodesics and distance fields.

A shortest geodesic between points $p_A$ and $p_B$ is defined by two distance fields, i.e., scalar fields of geodesic distances between the source points and all other points on the surface. Summing up the distance fields and tracing the isoline with the value $d(p_A, p_B)$ of the resulting distance field yields the connecting geodesic. Introducing an arbitrary third point $p = S(u,v)$ on the surface, we can state:
\begin{align}
    \label{eq:point_geodesic1} 
    d(p_A,p) + d(p,p_B) \ge d(p_A,p_B)\, ,
\end{align}
which is based on the triangle inequality of the surface metric. Expression~\eqref{eq:point_geodesic1} is an equality if and only if $p$ is on the geodesic connecting $p_A$ and $p_B$.

We now consider a second member between points $p_C$ and $p_D$ that intersects the first one. As above, we can state:
\begin{align}
    \label{eq:point_geodesic2} 
    d(p_C,p) + d(p,p_D) \ge d(p_C,p_D)\, .
\end{align}
Adding the distance fields emanating from $p_A,\,p_B,\,p_C$ and $p_D$, the minimum of the resulting scalar field takes the value 
\begin{align*}
d_{\min} = d(p_A,p_B)+d(p_C,p_D)\,. 
\end{align*}
Considering Expressions~\eqref{eq:point_geodesic1} and \eqref{eq:point_geodesic2}, this can only be the case if both inequalities become equalities and therefore the location of $d_{min}$ is a point on both shortest geodesics, i.e., the point of intersection.

In practice, we need to find the intersections on a mesh, where the distance fields are vectors that hold geodesic distances for all vertices. 
For every intersection, we first linearly interpolate the distance fields for the current $t$-values, superimpose them and find the minimum. Thus, the respective vertex is already close to the actual intersection of the geodesics. In a subsequent step, the result gets refined using second-order interpolation. To this end, we use the $(u,v)$-coordinates of the surface mesh parameterization of the vertex ring around the initial vertex, apply the values of the superimposed distance fields on a third axis, and fit a paraboloid. The $(u,v)$-coordinates of the minimum correspond to a point, which is sufficiently close to the real intersection of the two geodesics.

\revision{We reassemble the grid members using the appropriate intersections and increase their resolution by introducing extra points by linear interpolation between $(u,v)$-coordinates of adjacent intersections for sparse grids. }

\subsection{Non-Convex Boundaries}\label{sec:non-convex}

\begin{figure}
\centering
\includegraphics[width=\columnwidth]{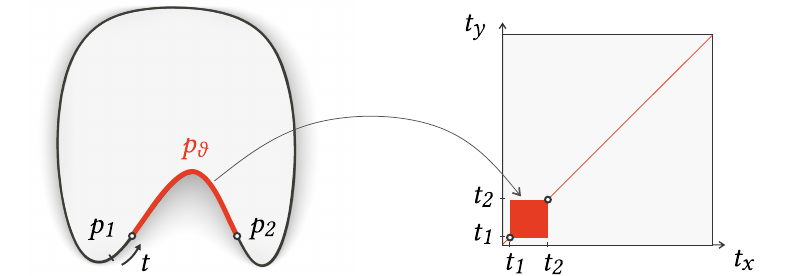}
\caption{Left: Non-geodesic shortest connections. 
We identify segments of the boundary that are non-convex using Equation~\eqref{eq:check_nc}. 
Right: All shortest connections along the red, non-convex segment of the boundary consti\-tute a square in the dual space ($p_1 = p(t_1)$, $p_2 = p(t_2)$). These curves are con\-sidered invalid for grid members. }
\label{fig:boundary_nc}
\end{figure}

Non-convex boundaries introduce curves $c(p_x,p_y)$ on the surface $S$ that minimize $d(p_x,p_y)$, however, are not shortest geodesics due to their non-vanishing geodesic curvature $\kappa_g$. Hence, we denote these curves as \textit{shortest connections}. Figure~\ref{fig:boundary_nc} depicts a shortest connection between the points $p_1$ and $p_2$ which runs along the boundary. We premise that the shape of the surface outside the boundary is unknown, i.e., a shortest geodesic between $p_1$ and $p_2$ cannot be found.

We call the boundary of a surface \textit{convex} if there are no \textit{shortest connections} that are tangential to the boundary, otherwise, we call it \textit{non-convex}. In the latter case, the parts of the boundary that make it non-convex need to be found. In Figure~\ref{fig:boundary_nc}, the lower bay of the boundary is such a non-convex part. 

Formally we identify non-convex boundaries by checking the following criterion: 
\begin{equation} 
    \label{eq:check_nc} 
    d(p_x,p_y) - d_b(p_x,p_y) = 0 \, ,
\end{equation}
where $d(p_x,p_y)$ is the shortest distance between $p_x,p_y$ on the surface and $d_b(p_x,p_y)$ is their distance measured along the boundary. If there is a combination of $p_x,p_y$  (with $t_x \neq t_y$)  and Equation~\eqref{eq:check_nc} is fulfilled, the boundary is non-convex and $c(p_x,p_y)$ leads entirely along the boundary. 
\revision{The part of the boundary that causes the non-convex behavior is formed by the points
$\pnc = p(\tnc)$ with $t_1 \le \tnc \le t_2$, where $t_1$ and $t_2$ are the smallest and largest $t$-values fulfilling Equation~\eqref{eq:check_nc}, respectively. }
In the dual space, such parts of the boundary appear as squares, please refer to Figure \ref{fig:boundary_nc} for a depiction. 

\begin{figure}
\centering
\includegraphics[width=\columnwidth]{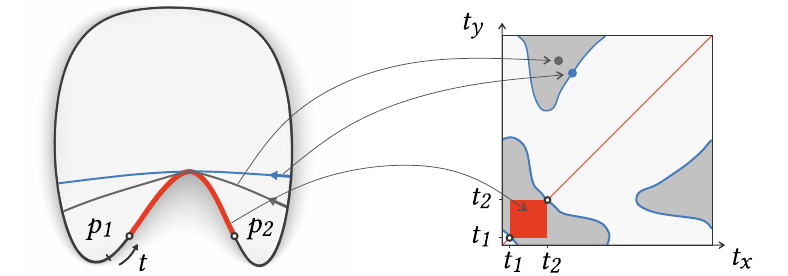}
\caption{
Left: Invalid members. The blue curve is tangential to the non-convex segment on the boundary, and the gray curve shares a part of it. 
Right: The blue curves emerge in the dual space on the boundary of the gray regions.
Curves in these regions are considered invalid and cannot be used as grid members. 
All such regions are detected using geodesic distances only. 
}
\label{fig:members_nc}
\end{figure}

If the boundary is non-convex, in a second step, we need to find all shortest connections $c(t_x,t_y)$ that lead partly along the boundary and exclude them from the set of valid grid members. This is important as, due to our initial assumption (cf.~Section~\ref{sec:background}), only lamellae based on geodesics ($\kappa_g = 0$) will bend properly to approximate the surface. Please refer to Figure~\ref{fig:members_nc}, where invalid regions are depicted in the dual space. 

\begin{figure}[b]
\centering
\includegraphics[width=\columnwidth]{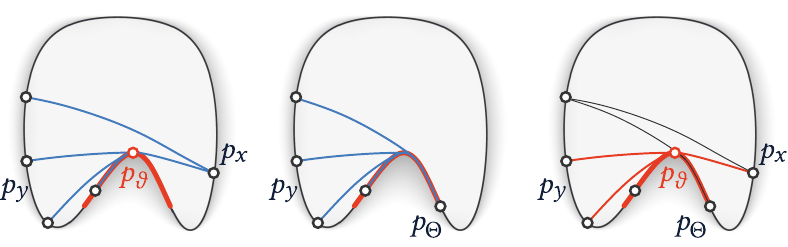}
\caption{
To identify all shortest connections emanating from $p_x$ to $p_y$ that go partially along the boundary, we proceed as follows:
Left: We compute the distances $d(p_x,p_y)$.
Middle: From $p_x$, we jump to the closest point $\pncc$ of the non-convex segment and compute the distances $d(\pncc,p_y)$.
Right: If shortest connections from $p_x$ to $p_y$ and $\pncc$ to $p_y$ have more than one common point, the shortest connection $c(t_x,t_y)$ is invalid.
}
\label{fig:finding_nc}
\end{figure}

To this end, we check the shortest connections from one fixed point $p_x$ to all other points $p_y$ on the boundary at a time. 
The resulting curves span a fan of shortest connections emanating from $p_x$ (cf. Figure~\ref{fig:finding_nc}).
We introduce a second fan emanating from $\pncc$ with the same destinations $p_y$. The point $\pncc$ is the point of $\pnc$ which is closest to $p_x$, measured along the boundary. 
We find all shortest connections that are no shortest geodesics by computing:
\begin{equation} 
    d_g(t_x,t_y) = d(p_x, p_y) - d(\pncc,p_y) \,,
    \label{eq:check_members} 
\end{equation}
and checking 
\begin{equation} 
    \frac{\partial d_g}{\partial t_y} = 0 \,.
    \label{eq:derivative_ty}
\end{equation}
\revision{
Equation~\eqref{eq:derivative_ty} is a necessary condition to find invalid connections $c(t_x,t_y)$ along the boundary. 
If $c(t_x,t_y)$ touches the boundary, we refer to the first point of contact as $p_\nc = p(\tnc)$. Assuming such a curve, we rewrite Equation~\eqref{eq:check_members}: 
\begin{equation*} 
\begin{aligned}[l,l]
    d_g(t_x,t_y) &= d(p_x, p_\nc) + d(p_\nc, p_y) -  d(\pncc,p_\nc) - d(p_\nc,p_y)\,, \\
    d_g(t_x,t_y) &= d(p_x, p_\nc) -  d(\pncc,p_\nc) \,,
\end{aligned}
\end{equation*}
$\nicefrac{\partial d_g}{\partial t_y}$ does indeed not depend on $t_y$ and vanishes.
Equation~\eqref{eq:derivative_ty} is not sufficient as 
there may be special cases where $d_g(t_x,t_y)$ remains constant w.r.t. changes in $p_y$ without boundary contact.
We have not encountered such cases in our models but can exclude them as they would appear as noise or isolated islands in the dual space, not connected to a non-convex segment (red square in Figure \ref{fig:boundary_nc}). 
}
Figure \ref{fig:finding_nc} helps to interpret Expression~\eqref{eq:check_members}: 
The difference $d_g(t_x,t_y)$ can be reduced to $d(p_x, p_\nc) -  d(\pncc,p_\nc)$, if the shortest connections from the fan of $p_x$ touch the boundary.

The set of invalid connections delivers in the dual space a map of connected regions that must be avoided in order to find valid geodesic grids. 

\subsection{Solution in the Discrete Domain}\label{sec:modify_combinatorics}

For the computation of invalid regions, we resort to the discrete domain $\mathbb{N}$ and introduce a mapping $z$ which maps all $t$-values to integer coordinates of the mesh boundary: 
\begin{align*}
z\colon t\in [0,1) \mapsto i\in [1,\dots, N] \,, 
\end{align*}
where $N$ is the mesh boundary resolution. We denote the coordinates of the curves in the discrete space as $c(i_x,i_y) = c(z(t_x), z(t_y))$, where $i_x,i_y \in [1, \dots, N]$ represent their counterparts $t_x,t_y \in [0,1)$ from the continuous domain. Note that $N+1 = 1$ and $c(i_x,i_y) = c(i_y,i_x)$, respecting the symmetry given by $\ttrans$. 

\begin{figure*}[t]
\centering
\includegraphics[width=\textwidth]{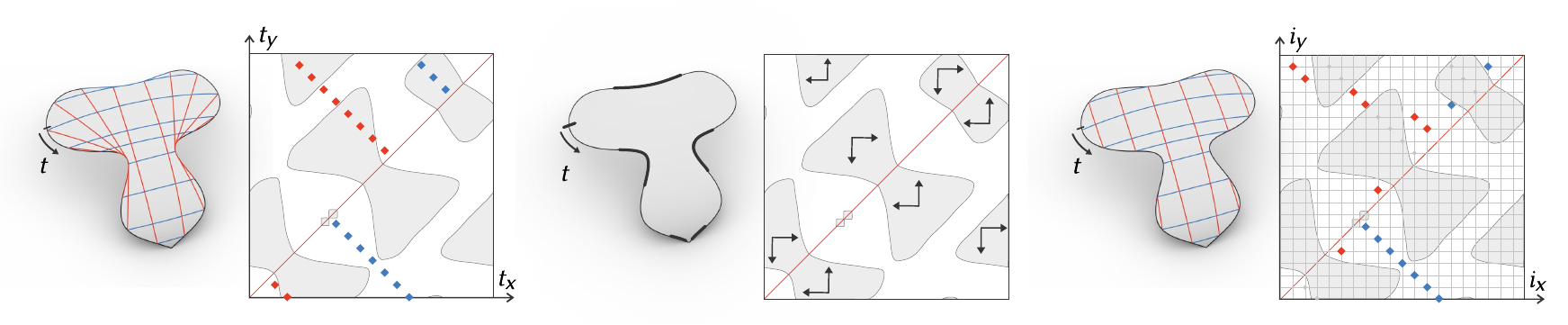}
\caption{
Layout correction algorithm.
Left: An invalid grid, members appear as points in the dual space. Middle: For every invalid region in the dual space, there are fixed shifting directions that govern the adjustment of layouts.
Right: Members shifted out of invalid regions to the next feasible combination in the discrete  dual space $z\colon(t_x,t_y)\mapsto(i_x,i_y)$. Note, it only changes the combinatorics of the grid but leaves the endpoint locations unchanged. 
The small square-shaped regions in the dual space correspond to two small, almost straight, non-convex segments of the boundary at the lower tip of the surface.
Please recall that the dual space and the invalid regions are periodic, as depicted in Figure \ref{fig:tspace}. }
\label{fig:member_correction}
\end{figure*}

To initially find non-convex regions we compute Equation~\eqref{eq:check_nc} on the mesh. We reuse the distance fields and perform:
\begin{equation} 
    \label{eq:check_nc_mesh} 
    D - B < \epsilon \, ,
\end{equation}
where $D$ and $B$ are $N$x$N$ matrices, $D$ is the distance map $D(i_x,i_y) = d(p(i_x),p(i_y))$, $B$ holds the distances measured along the boundary, $B(i_x,i_y) = d_b(p(i_x),p(i_y))$,
and $\epsilon$ is a small constant close to zero. The distances $d(p_x, p_y)$ and $d(\pncc, p_y)$ in Equation~\eqref{eq:check_members} are a subset of the distances from the precomputed distance fields, the partial derivative w.r.t. $t_y$ becomes a difference of $d_g(i_x,i_y)$, w.r.t. $i_y$.

Now we can formulate the search for invalid regions using a region growing approach as in image processing. We know the core regions from the difference in Expression \eqref{eq:check_nc_mesh}, so  we grow them along respective $i_x$ and $i_y$ coordinates as
\begin{align*}
    c(i_x,i_y) = c(i_x + s_{i_x}, i_y) \, \\
    c(i_x,i_y) = c(i_x, i_y + s_{i_y}) \,
\end{align*}
where $s_{i_x}$ and $s_{i_y}$ 
are steps that are either $+1$ or $-1$ depending on which half-space of the dual space the points are located. During this movement, distance checks as in Equation~\eqref{eq:check_members} are performed by looking up distance values in the distance map $D(i_x,i_y)$. Performing the procedure until convergence yields masks that contain invalid regions. 

During the optimization of grids (cf. Section~\ref{sec:model:optim}), grid members need to be found frequently, and the combinatorics of the grid has to change in order to produce valid grid members. 
\revision{
We use the dual space and its invalid regions, identified using Equation \eqref{eq:derivative_ty}, to correct layouts efficiently. 
Figure \ref{fig:member_correction} depicts how invalid members are detected and the subsequent adjustment of a layout.} 
In cases where a member arrives in an invalid region, the respective member family ($g$ or $h$) is split into two along their respective $i_x$ and $i_y$ coordinates in the dual space, and all following members are shifted out until the invalid-region condition as well as grid consistency constraints, which maintain the order of elements, are fulfilled. 

This algorithm can be implemented very efficiently in the discrete domain, where look-ups in the precomputed distance map and invalid-region maps are performed.

\breakpage
\section{Elastic Geodesic Grid Energy}\label{sec:grid_energy}

In this section, we describe the mathematical model of our approach, based on insights from differential geometry and variational principles, and derive the necessary energy functionals which constitute our elastic geodesic grids. 

\begin{figure}[b]
\centering
\includegraphics[width=\columnwidth]{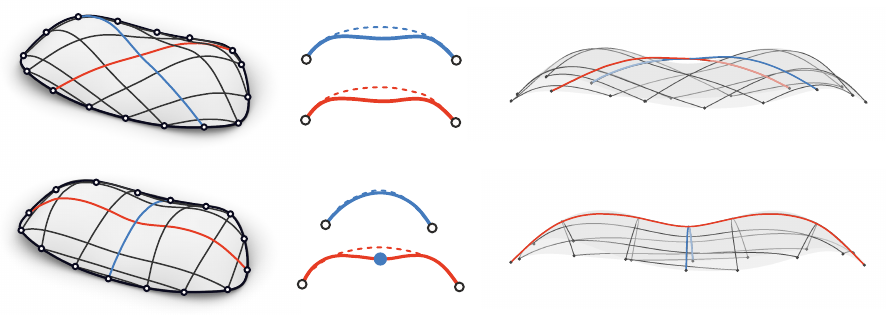}
\caption{Left: Two different grid layouts on the same surface. Middle (top): The blue and red geodesics are prone to relax to different shapes. Middle (bottom): The red geodesic is constrained by the blue one. Right: Physical simulation of the grids, only the lower grid nestles to the surface well.}
\label{fig:stable_layout}
\end{figure}

The main aspect of our work is finding networks of intersecting curves that nestle to the surface when boundary conditions are enforced, i.e., when grids are fixed to anchors. However, not all such grids will both nestle to the surface well and additionally also closely capture the surface's features, like local extrema of $K$. 
Figure~\ref{fig:stable_layout} shows two different geodesic grids and their respective equilibrium shapes when boundary conditions are enforced. 
As evident, some grids are better suitable than others and their choice obviously depends on local surface features. 

\subsection{Elastic Geodesic Curves}\label{sec:model:curves}

The internal bending energy of an arc length parameterized curve $c \coloneqq \mb{c}(s)$ of length $l$ can be formulated using the Elastica energy:
\begin{equation}
E_b = \int_0^l \kappa^2 \,\dd s  \,, 
\label{eq:bending_1}  
\end{equation}
where $\kappa = \norm{\mb{c}''(s)}$ is its curvature, i.e., the length of the curvature vector. 

We can use this energy to approximate the bending behavior of the centerlines of slender wooden lamellae quite accurately. 
This assumption is not entirely true in theory for non-stretchable, perfectly developable strips, which indeed need to maintain the same length of the centerline as well as of the edges. 

Nonetheless, wooden lamellae are not perfectly developable, and hence their behavior is more similar to the bi-normal strips \cite{Wallner2010}, where the ratio of the lengths of the center curve and the edges varies slightly if they are bent and twisted (i.e., the strip stretches or compresses slightly). However, the deviation is negligible if the width to length ratio of the strip is small, as in our case. 
This assumption is also common in physical models, like discrete elastic rods~\cite{Bergou2008}, which allow for the simulation of slender rods with varying cross-sectional ratios. 
For these reasons, we choose to approximate the strips with their center lines. 

For curves $c(s)$ which lie on a surface $S$, their curvature vector $\mb{c''}$ can be decomposed into its normal curvature component $\kappa_n$ and geodesic curvature component $\kappa_g$, such that it is given by
\[
\mb{c}^{\prime\prime} = \kappa_n \mb{n}_S + \kappa_g \mb{b}_c \,,
\]
where $\mb{n}_S$ is the surface unit normal and $\mb{b}_c$ is a unit vector in the tangent plane, orthogonal to the curve tangent. This dependency can also be expressed by the Pythagorean theorem as: 
\[
\kappa^2 = \kappa_n^2 + \kappa_g^2 \,.
\]

As per definition (cf.~Section~\ref{sec:background}), geodesic curves have vanishing geodesic curvature, i.e., $\kappa_g=0$, the curve's normal is aligned with the surface normal $\mb{n}_S$.
Since we allow only geodesic curves on $S$, the bending energy in Equation~\eqref{eq:bending_1} reduces to 
\begin{equation}
E_b = \int_0^l \kappa_n^2 \, \dd s \,. 
\label{eq:bending_2} 
\end{equation}
We can further rewrite $\kappa_n$ using Euler's theorem as
\[
\kappa_n = \kappa_1\,\cos^2\varphi + \kappa_2\,\sin^2\varphi \,,
\]
where $\kappa_1$, $\kappa_2$ are the principal curvatures at the given surface location, and $\varphi$ is the angle between the curve's tangent vector $\mb{t}_c$ and the principal curvature direction $\mb{v}_1$. Assuming a unit $\mb{t}_c$, it can also be expressed in the surface's local frame spanned by the principal curvature directions $\mb{v}_1$, $\mb{v}_2$ in terms of inner products as
\[
\kappa_n =  (\kappa_1-\kappa_2)\inner{\mb{t}_c}{\mb{v}_1}^2  + \kappa_2 \,.
\]

To evaluate $\kappa_n$ at vertices $q(u,v)$ of grid members, we approximate $\mb{t}_c$ using the normalized  mean of their adjacent edges and linearly interpolate values of $\kappa_1,\kappa_2$, and $\mb{v}_1$ for the respective $(u,v)$-coordinates.


\subsection{Shape Stability}
\label{sec:shape_stability}

\begin{figure}
\centering
\includegraphics[width=\columnwidth]{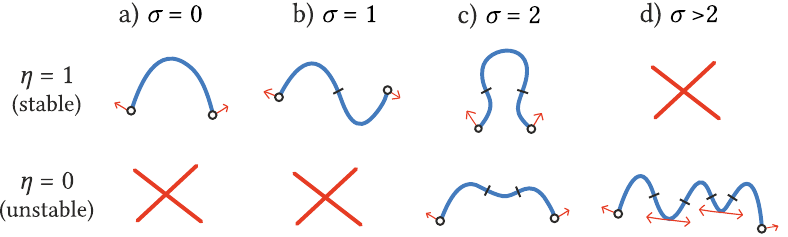}
\caption{\revision{We estimate the shape stability of geodesics using the shape stability parameter $\eta$, which depends on the number $\sigma$ of inflection points w.r.t. the normal curvature $\kappa_n$. We expect that all curves of type d) need additional support from crossing members. Curves of type c) are a limit case and are assessed using the criterion of \cite{hafner2021}.} }
\label{fig:stable_unstable}
\end{figure}

\begin{figure}
    \includegraphics[width=\columnwidth]{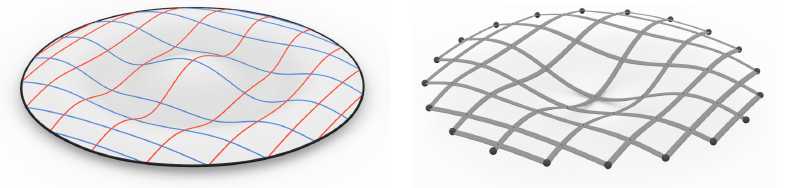}
	\caption{\revision{Failure case. Left: A desired shape and geodesic grid. Right: The result of physical simulation. Due to the inner bump and the symmetry of the surface, no stable shape can be determined. As not enough stable members can be found, the physical grid does not follow the desired shape.  }  }
	\label{fig:failure_example}
\end{figure}

\revision{
The shape of a physical grid depends on several factors, like the bending energy of grid members and mutual stabilization. To ensure a close approximation of the target surface but avoid costly physical simulation, we introduce an energy that estimates how closely a grid will preserve its initial shape when boundary conditions are enforced. 
To define this energy, we first analyze the quality of members individually, then account for mutual stabilization of members, and finally obtain a single scalar value $E_{\msub{shape}}$ ranging between 0 and 1. The lower its value, the larger the deviation from the target surface can be expected.

\begin{figure}
\centering
\includegraphics[width=\columnwidth]{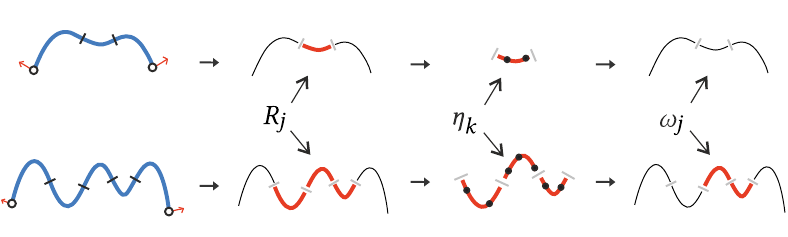}
\caption{\revision{We account for mutual stabilization effects by dividing the curve into segments $R_j$ needing support and checking if they are stabilized by crossing members (black dots),
taking into account their shape stability $\eta_k$.
Top: The central region of the curve is stabilized, $\eta = 1$. Bottom: Only one of three curve segments is stabilized, we expect poor performance, $\eta = \nicefrac{1}{3}$. }  
}
\label{fig:regions_unstable}
\end{figure}

We initially assess the quality of grid members individually 
by means of the number $\sigma$ of inflection points of each member w.r.t. the normal curvature $\kappa_n$. We consider members with $\sigma < 2 $ as stable and members with $\sigma > 2 $ as unstable, please refer to Figure \ref{fig:stable_unstable} for a depiction. To classify members with $\sigma=2$, we use a stability criterion for planar curves \cite{hafner2021} (neglecting the torsion of grid members).
We subsequently classify each curve with a shape stability parameter $\eta = 0 \text{ or } \eta = 1$.

The interaction of elastic members in a grid is highly complex, however, we want to take into account mutual stabilization effects for unstable members ($\eta = 0$). To this end, we divide them into $(\sigma-1)$ curve segments $R_j$  needing external support, which are separated by the inflection points (cf. Figure \ref{fig:regions_unstable}). For each segment $j$, we check for stable crossing members and assign a stability weight $\omega_j$:
\begin{equation*}
    \omega_j =  
    \begin{cases}
    0 \quad\text{if}\quad \normalsize\sum_{k=1}^{n+m} B(\eta_k \in R_j) \eta_k \ge 1 \,, \\
    1 \quad\text{if}\quad \sum_{k=1}^{n+m} B(\eta_k \in R_j) \eta_k  = 0 \,,\\
    \end{cases}
\end{equation*}
where $B(\cdot)\to\{0,1\}$ is a Boolean operator which ensures that only $\eta$ of crossing members intersecting in the respective curve segment are summed up, and $k$ iterates over all members. If one or more curves with $\eta = 1$ cross in a segment needing support, it is considered as stabilized, and $\omega_j = 1$, otherwise $\omega_j = 0$.
We subsequently update $\eta$ of all unstable members by the arithmetic mean of their $\omega$-values: 
\begin{equation*}
    \eta = \frac{1}{\sigma-1}\sum_{j=1}^{\sigma-1} \omega_j \,,
\end{equation*}
and finally assess the shape quality of the whole grid by taking the shape stability parameters of all members into account:
\begin{equation}
    E_{\msub{shape}} = \frac{1}{n+m}\sum_{k=1}^{n+m} \eta_k \,.
\end{equation}

Figure \ref{fig:failure_example} shows a failure example with a ripple-like target surface. The grid reaches a low shape stability ($E_{\msub{shape}}=0.45$) which indicates that we can expect a poor approximation of the surface. }

\subsection{Least and Most Effort Energy}\label{sec:model:grids} \label{sec:bending:energy}

Our goal is to approximate arbitrary representable surfaces, hence we need an energy functional that allows us to 
\revision{ capture the surface characteristics well.
To ensure this, we define two types of curves, which must be present in a grid.

Among all geodesics on $S$, the subset minimizing the bending energy in Equation~\eqref{eq:bending_2} is most likely to produce lamellae that nestle to the surface well, as shown in  Figure \ref{fig:stable_layout}, bottom. Hence, such minimal energy geodesics are essential candidates for our grids. On the other hand, we also require curves that capture high-$K$ bumps on the surface.

We denote the first kind as} \textit{least-effort} geodesics; 
they run along hyperbolic regions, between elliptic regions (bumps) on the surface. They minimize their arc length and their curvature is low.
The second kind are \textit{most-effort} geodesics; they maximize $\kappa_n^2$ along their trajectory and their arc length. At the same time, they are attracted to the geometrically exposed features of a surface; their curvature is high and potentially oscillates, as they may travel over multiple elliptic regions. 
To identify such curves, we formulate their \textit{effort energy} as the integral of the squared normal curvature along the curve $c$, normalized by the curve length: 
\begin{align}
    E_c(c(t_x,t_y)) = \frac{1}{l_{c}} \int_{c} \kappa_n(u(s),v(s))^2 \,\dd s \,.
\label{eq:effort_curve}
\end{align}

Next, we extend this formulation to the entire grid, such that it ensures that we get as many \emph{least-effort} geodesics and as many \emph{most-effort} geodesics as possible at the same time. As $E_c$ are the energies of our grid members, this is equivalent to pushing them apart as far as possible. Hence, the energy we use to find suitable grid configurations is
\begin{align}
    E_{\msub{effort}} = - \sum_{i=1}^{n+m} \pare{ E_{c,i} - \widehat{E}_c }^2  \,, 
\label{eq:grid}
\end{align}
where  
\[
\widehat{E}_c =  \frac{1}{n+m}\sum_{i=1}^{n+m} E_{c,i} \, 
\]
is the arithmetic mean of the member energies $E_c$, and $(n+m)$ is the number of all member curves in the grid. The rationale behind Equation~\eqref{eq:grid} is that we maximize the variance of the effort energies, and thus curves are forced to either minimize or maximize their value of $E_c$. 

In practice, we discretize the elastic energy $E_c$ of each geodesic by a weighted sum of $\kappa_n^2$, evaluated on the surface mesh, using its parameterization at \revision{curve points $q(u,v)$}:
\begin{equation*}
    E_c = \frac{1}{l_{c}} \sum_j \frac{1}{2}\left|\left|(q_{j}-q_{j-1})\right|\right|\, \kappa_n(q_j)^2 +\frac{1}{2}\left|\left|(q_{j+1}-q_{j})\right|\right|\, \kappa_n(q_j)^2 \,,
\end{equation*}
where $j$ iterates over all curve points of a given member curve $c$.

\subsection{Minimization of Elastic Grid Energy}\label{sec:model:optim}

To find the best orientation for the whole grid, we minimize the energies proposed in Sections~\ref{sec:shape_stability} and \ref{sec:bending:energy} using the following objective: 
\begin{equation}\label{eq:e_grid}
    E_{\msub{grid}}(\mb{t}) = E_{\msub{effort}}(\mb{t}) - \lambda \, E_{\msub{shape}}(\mb{t}) \,,
\end{equation}
where $\lambda$ is a weighting factor.
This functional is \revision{a piecewise smooth energy, however, we must expect jumps due to the changes in combinatorics} and multiple local minima, which occur when a subset of members has a beneficial orientation, but the rest of the grid has not.

As the grid is parameterized by the locations of the anchor points on the boundary, the variables in our problem are $\mb{t}=[t_1, \dots, t_k]$. The grid combinatorics are unchanging in case the surface has a convex boundary, but they need to be adapted in every iteration in case the boundary is non-convex (cf. Section \ref{sec:modify_combinatorics}). This can change the number of anchor points, making it particularly difficult to solve with a gradient-based continuous optimization approach. 
To address this issue, we use a genetic algorithm~\cite{Goldberg1989} to find the best orientation for the whole grid. 
We formulate the optimization problem as 
\revision{
\begin{gather}
\min_{\mb{t}}\, E_{\msub{grid}}(\mb{t}) \; \st \, 
\begin{cases}
    \; \forall\; t_1,\dots,t_k \in [0,2) \\
    \; \mb{A}\mb{t} \leq \mb{b} 
\end{cases}.
\label{eq:optimization} 
\end{gather}
}
In each optimization step, we allow the values of $\mb{t}$ range  $\in [0,2)$ in order to deal with the seam of the boundary curve. Subsequently, they are transformed back to $[0,1)$ respecting the symmetry given by the transformation $\ttrans$ in every iteration. To avoid leapfrogging, we apply linear inequality constraints that secure monotonously growing entries in $\mb{t}$ and a certain minimum distance between member endpoints. 

We further provide the GA with an initial population, which is equivalent to rotating the grid on the surface in a number of steps. We observed that this step already roughly sets the orientation of the grid and speeds up convergence.

Our grid design algorithm is implemented in \matlab, and we use its GA-solver for solving the Optimization Problem \eqref{eq:optimization}.


\section{Grid Planarization}\label{sec:grid_planarization}

Arbitrary geodesic grids cannot be transformed to a planar state without changing either the lengths of members or the locations of their inner intersections. 
To planarize the grid, we resolve to the second solution and use the concept of sliding notches~\cite{Pillwein2020}. 
This type of connection allows for a short amount of sliding at the connection of members and thus provides two additional translational degrees of freedom at each connection.

\subsection{Planarization Algorithm}\label{sec:grid_planarization:planarization}
\revision{
We express the locations of inner intersections w.r.t. the barycentric coordinates $\mathbf{\lambda}, \mathbf{\tod{\lambda}}$ on the members, where overlined quantities refer to the planar configuration (cf. Figure \ref{fig:planarization_overview}).
For an intersection $q,\tod{q}$, the coordinates are:
\begin{equation*}
    \begin{aligned}
    &&\lambda_g = \frac{d(p_A,q)}{l_g}\,  && \tod{\lambda}_g = \frac{d(\tod{p}_A,\tod{q})}{l_g}\, && l_g = d(p_A,p_B) = d(\tod{p}_A,\tod{p}_B)\,,\\
    &&\lambda_h = \frac{d(p_C,q)}{l_h}\,  && \tod{\lambda}_h = \frac{d(\tod{p}_C,\tod{q})}{l_h}\, && l_h = d(p_C,p_D) = d(\tod{p}_C,\tod{p}_D) \,.
    \end{aligned}
\end{equation*}

\begin{figure}
\centering
\includegraphics[width=\columnwidth]{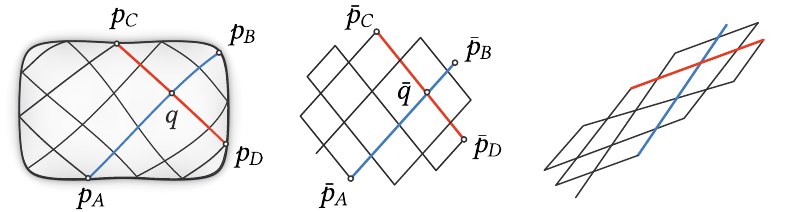}
\caption{\revision{Left: For intersection $q$, we determine partial lengths $d(p_{A},q)$, $d(p_{C},q)$, and total lengths $d(p_{A},p_{B})$, $d(p_{C},p_{D})$.  Middle: We initialize the planar grid, and parameterize it by its endpoint coordinates. Matching partial and total lengths with the corresponding values on the surface is generally not possible. Right: Our planarization algorithm maintains total lengths perfectly and partial lengths as well as possible.} }
\label{fig:planarization_overview}
\end{figure}

We planarize the grid by first generating an initial planar grid of the same combinatorics and formulate a quadratic optimization problem w.r.t. its 2d endpoint coordinates. Therefore we compute the points 
\begin{equation*}
    \begin{aligned} %
    &\tod{q}_g = \lambda_{g} \, \tod{p}_A + (1-\lambda_g) \, \tod{p}_B \,, \\ 
    &\tod{q}_h = \lambda_{h} \, \tod{p}_C  + (1-\lambda_h) \, \tod{p}_D\,, \\
    \end{aligned}
\end{equation*}
which are defined w.r.t.  $\lambda_g, \lambda_h$ of the geodesic grid and ideally should coincide with $\tod{q}$, please refer to Figure \ref{fig:planarization_q} for a depiction. 
\begin{figure}[b]
    \begin{minipage}[c]{0.62\columnwidth}
        \caption{\revision{In order to keep our optimization problem quadratic, we do not explicitly compute $\tod{q}$, but use the barycentric coordinates $\lambda_g,\lambda_h$ to compute $ \tod{q}_g,\tod{q}_h $. If $\lambda_g =\tod{\lambda}_g$ and $\lambda_h =\tod{\lambda}_h$, then $ \tod{q}_g = \tod{q}_h = \tod{q} $ and the connection has no notch, otherwise we minimize the distance between $ \tod{q}_g$ and $\tod{q}_h$.} }
        \label{fig:planarization_q}
    \end{minipage}
    \hfill
    \begin{minipage}[c]{0.32\columnwidth}
        \vspace{0.1cm}
        \includegraphics[width=\textwidth]{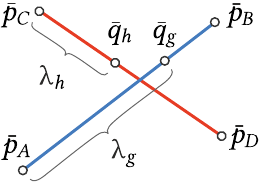}
    \end{minipage}
\end{figure}
}
Since we want to minimize the length of the notches, we formulate the planarization problem as a constrained quadratic minimization with the following objective: 
\revision{
\begin{equation}
    E_{\msub{notch}} = \sum_{j=1}^{n_q} \Vert\, \tod{q}_{g_j}-\tod{q}_{h_j} \Vert^2 \,,
\end{equation}
where $j = [1 \dots n_q]$ denotes the intersections of the grid. 
}

To ensure that the planar grid elements perfectly maintain the total lengths of their spatial counterparts, we introduce hard constraints which enforce that lengths between curves on the surface and in the plane match:
\revision{
\begin{align*}
    G_{\msub{len}} = \Vert\,\tod{p}_y-\tod{p}_x\,\Vert^2 -  d({p}_y,{p}_x)^2 = 0 \,, 
\end{align*}
}
where $\tod{p}_y,\tod{p}_x$ and ${p}_y,{p}_x$ are the endpoints of corresponding members in the plane and on the surface.

\subsection{Fabrication Constraints}\label{sec:grid_planarization:fabrication}
\begin{figure}
\centering
\includegraphics[width=\columnwidth]{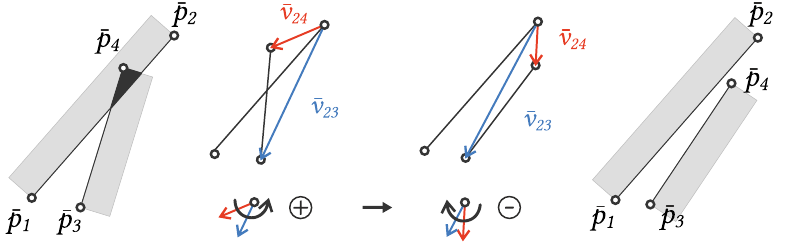} 
\caption{\revision{Endpoints $\tod{p}_1, \tod{p}_2, \tod{p}_3, \tod{p}_4$ of two consecutive lamellae are defined by the offset $\nicefrac{w}{2}$ from the centerline. We avoid overlapping by ensuring that certain determinants of vectors connecting the endpoints have the right sign.} 
}
\label{fig:fab_constraints}
\end{figure}

For manufacturing purposes, the width $w$ of lamellae must be considered. Therefore we add optional minimum distance constraints, ensuring an offset between consecutive members of the same family.
To this end, we use the endpoints $\tod{p}_{1}, \tod{p}_{2}, \tod{p}_{3}, \tod{p}_{4}$ of two consecutive members of a family (cf. Figure~\ref{fig:fab_constraints}). We introduce the constraints w.r.t. the sign of determinants of vectors between the endpoints:
\begin{align*}
    e_1 &= \det\pmat{\tod{\mathit{v}}_{\,14},\tod{\mathit{v}}_{\,13}} \leq 0 \,,\\
    e_2 &= \det\pmat{\tod{\mathit{v}}_{\,24},\tod{\mathit{v}}_{\,23}} \leq 0 \,,\\
    e_3 &= \det\pmat{\tod{\mathit{v}}_{\,31},\tod{\mathit{v}}_{\,32}} \leq 0 \,, \\
    e_4 &= \det\pmat{\tod{\mathit{v}}_{\,41},\tod{\mathit{v}}_{\,42}} \leq 0 \,,
\end{align*}
where each $e_i$ checks one endpoint and positive values indicate overlaps. Since these constraints are rather restrictive, we introduce them as soft constraints within the objective as:
\begin{align*}
E_{\msub{fab}} = \sum_{k=1}^{n-1} \sum_{i=1}^4 B(e_{i,k}>0)(e_{i,k}) + \sum_{k=1}^{m-1} \sum_{i=1}^4 B(e_{i,k}>0)(e_{i,k}) \,,
\end{align*}
where $B(\cdot)\rightarrow\{0,1\}$ is a Boolean operator which is used to discard cases where the constraints are not violated and $k,i$ iterate over all pairs of neighboring grid members $g=\{c_1,\dots c_n\}$ (fist sum) and $h=\{c_1,\dots c_m\}$ (second sum). 

We minimize the planarization objective using Pareto weighting of the notch and fabrication terms
\begin{align}
    E_{\msub{planar}} = E_{\msub{notch}} + \mu E_{\msub{fab}} \quad \st \quad G_{\msub{len}}=\mathbf{0} \,,
    \label{eq:planarization}
\end{align}
Please note that the objective function and the constraints are quadratic w.r.t. the variables and an analytic gradient can be computed. We solve this optimization problem using sequential quadratic programming in \matlab. 

\breakpage

\section{Results and Evaluation}\label{sec:results}

\begin{figure*}
    \centering
    \includegraphics[width=\textwidth, trim=0 0 0 0, clip]{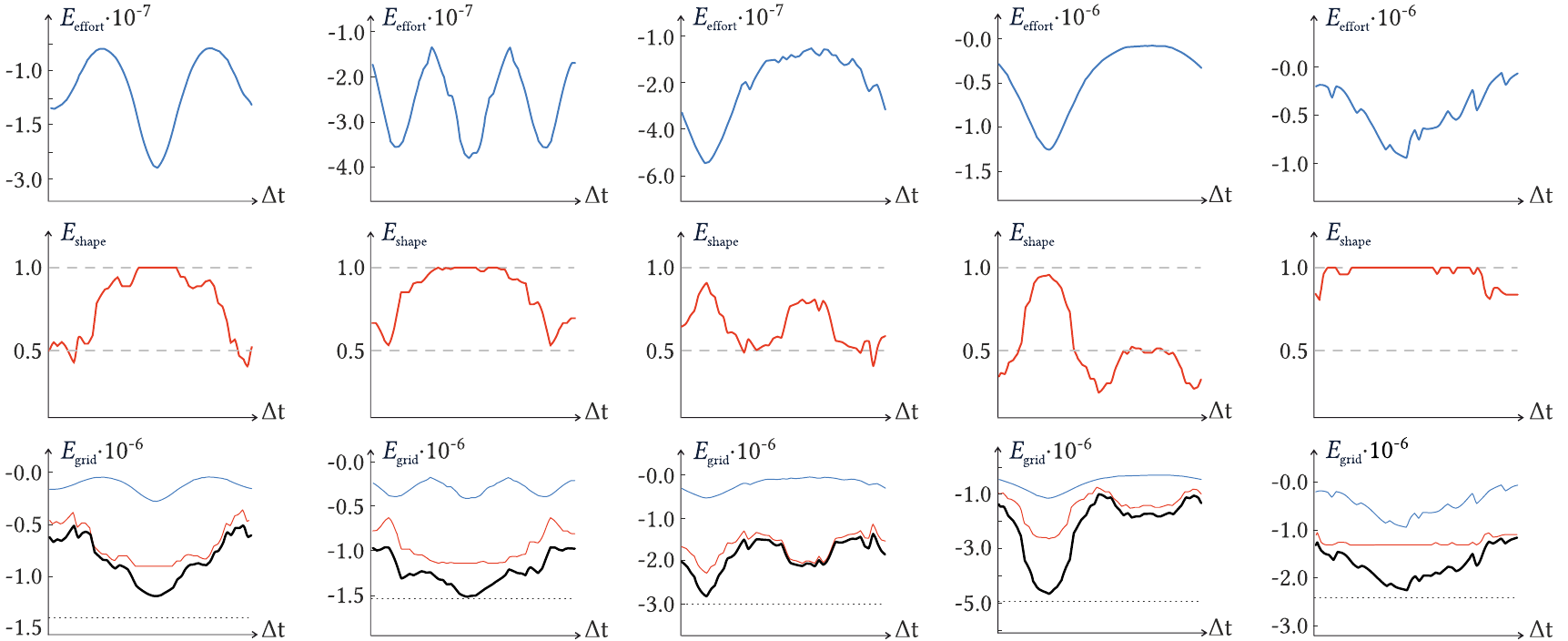}
    \caption{
    \revision{Energies $E_{\msub{effort}}$, $E_{\msub{shape}}$ and $E_{\msub{grid}}$ for our fabricated examples of Figures \ref{fig:results} and \ref{fig:results_additional}. The plots show the energies when the grid is rotated on the surface for a quarter rotation; the black dotted lines represent the final energy of the grid after optimization. Jumps in the energies indicate changes in combinatorics in the grid (cf. Equation ~\eqref{eq:e_grid}). From left to right: Starship, Bumps, Flower, Hills, and Moon.} }
    \label{fig:effort_plots}
\end{figure*}

Using our method, we have represented a number of surfaces which are depicted in Figures~\ref{fig:results} and \ref{fig:results_additional}. The input surfaces have positive and negative Gaussian curvature regions, inner bumps, convex and non-convex boundaries. 

We fabricated the models from lime wood or acrylic glass, the lamellae were laser-cut from thin plates, connected by screws, and placed on 3d-printed supports after assembly. The supports have inclined contact areas to enforce the desired shape of the grid (cf. Section \ref{sec:objectives}). 

\subsection{Quantitative Results}

\begin{figure*}
\centering
\includegraphics[width=\textwidth, draft=false]{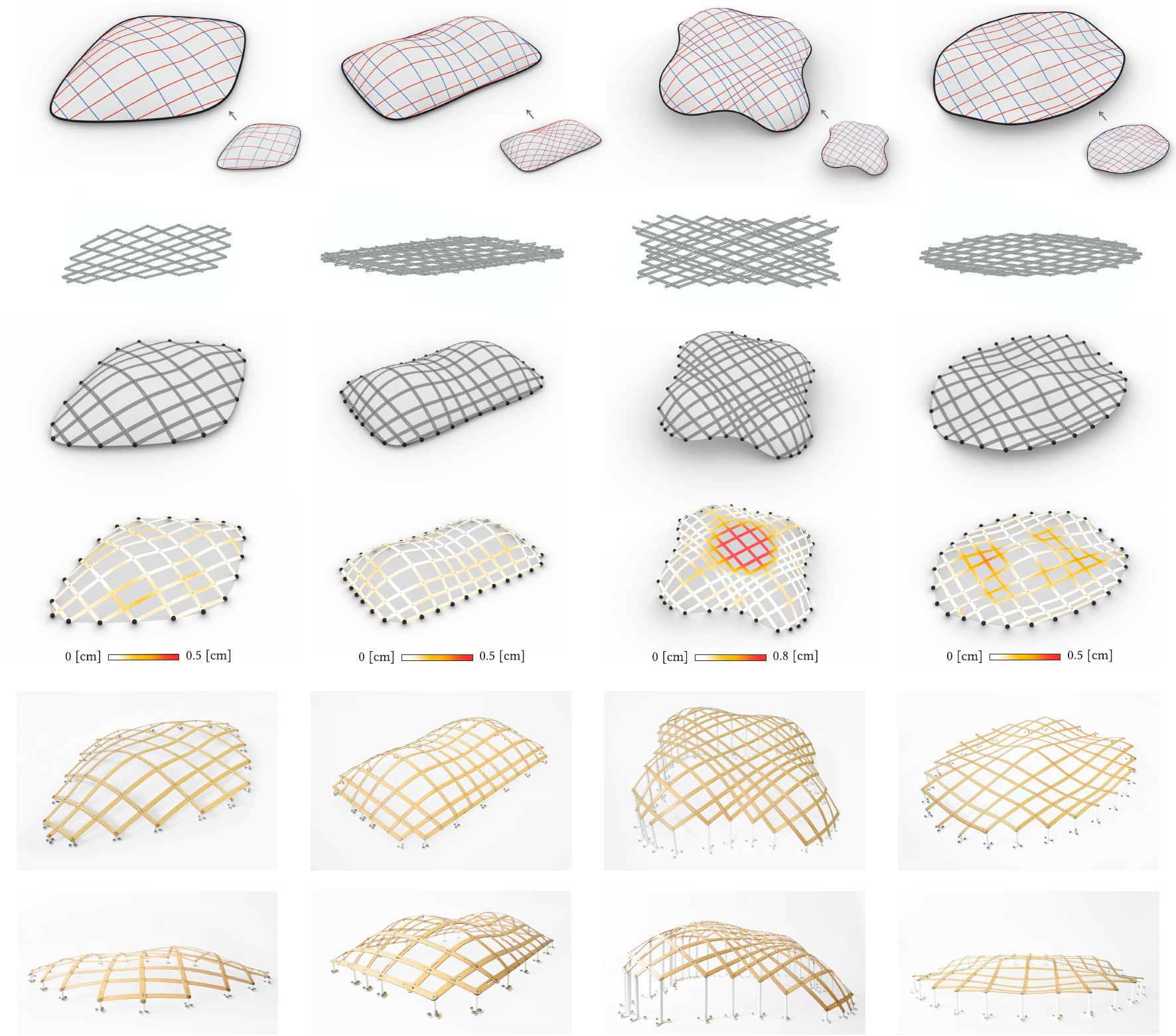}
\caption{Computed, simulated, and fabricated results of our method.  From left to right: Starship, Bumps, Flower, Hills. 
All grids nestle the respective surfaces well and capture their characteristics. Example Flower is a limit case: The central inward bump cannot be captured by an elastic gridshell without inner supports (cf. Section \ref{sec:discussion}). However, example Hills shows that inner bumps can be feasible without inner supports if enough \emph{least-effort} and \emph{most-effort} geodesics can be found. When it is deployed and fixed to 3d-printed anchors with inclined contact areas, the shape of a grid emerges.
\revision{ Please note that the fabricated results were optimized using a previous version of energy $E_{\msub{effort}}$ and were not re-fabricated because the deviations are not noticeable in the models.} Best seen in the electronic version in closeup.}
\label{fig:results}
\end{figure*}

\begin{figure*}
\centering
\includegraphics[width=\textwidth,  draft=false]{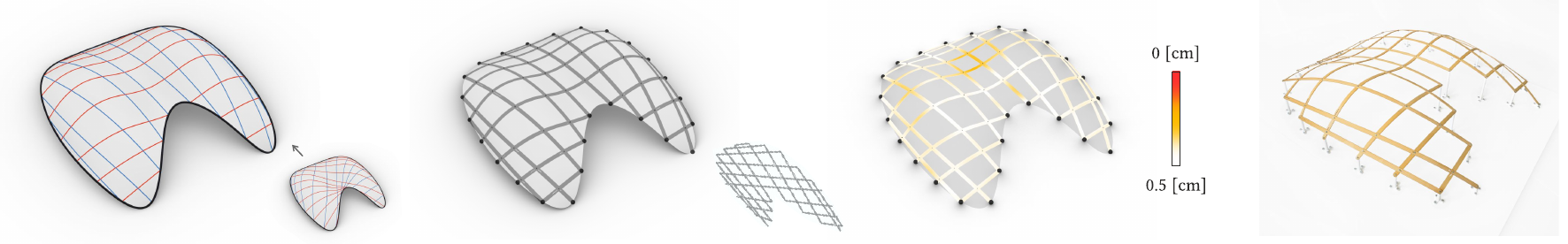}
\caption{Computed, simulated, and fabricated example Moon.
The red family of geodesics is split into sub-families due to the non-convexity of the boundary. The physical grid nestles the surface very well. \revision{Please note that the fabricated result was optimized using a previous version of energy $E_{\msub{effort}}$ and was not re-fabricated because the deviations are not noticeable in the model.}}
\label{fig:results_additional}
\end{figure*}

In Table \ref{tab:commands} we summarize quantitative results of our method for the models depicted in Figures \ref{fig:results}, \ref{fig:results_additional}, and~\ref{fig:failure_example}. To check the agreement of the grid shape and the target surface, we simulated the physical behavior of the deployed grid using the discrete elastic rods model \cite{Bergou2008}. We refer the reader to the paper for details. In Table \ref{tab:commands},  $\text{RMS} ~\Delta$ denotes the root mean square distance between grid vertices and the base mesh and $\text{max} ~\Delta$ is the maximum distance.

\begin{table}[b]
	\small
	\caption{
	\revision{
	Quantitative results of our method. We measure the root mean square error ($\text{RMS}~\Delta$) and the maximum error ($\text{max} ~\Delta$) between the member centerlines and the target mesh in centimeters; $\varepsilon_{\text{geo}}$ refers to the mean deviation between traced geodesics and our reconstruction w.r.t. the mean edge-length in percent.
	Timings are in seconds,  $t_{\text{pre}}$ refers to computation of distance fields and analysis of the boundary, $t_{\text{opt}}$ refers to the GA-convergence times, and $t_{\text{pla}}$ to the planarization of the grid. $|M_V|$ expresses the number of mesh vertices and $n$ the number of grid members.
	Measured on an AMD Ryzen 7 1700 Eight-Core using parallel computing.
	} }
	\label{tab:commands}
\begin{tabular}{r r r r r r r}
	& \multicolumn{1}{c}{Starship} & \multicolumn{1}{c}{Bumps} & \multicolumn{1}{c}{Flower} &  \multicolumn{1}{c}{Hills}& \multicolumn{1}{c}{Moon} & \multicolumn{1}{c}{Drop}   \\
	\midrule
	width 	    &60.0&  60.0 &  55.5 &  60.0 &   54.2 &  \revision{60.0}     \\
	depth 	    &40.4 &  37.5  & 60.0 &	 47.5 &   60.0 & \revision{60.0}   \\
	height 	    &6.8&	 8.9 &   20.0  & 5.9 &  10.6 & \revision{4.2}	  \\[0.1cm]
	
	$|M_V|$   &    \multicolumn{1}{r}{3734}&    \multicolumn{1}{r}{3333}  &    \multicolumn{1}{r}{3975} &	  \multicolumn{1}{r}{ 5102}& \multicolumn{1}{r}{ 3437}& \multicolumn{1}{r}{\revision{4591}} \\

	$n$   &    \multicolumn{1}{r}{18}	&    \multicolumn{1}{r}{22}  &   \multicolumn{1}{r}{30} &   	  \multicolumn{1}{r}{22 }& \multicolumn{1}{r}{ 22 }& \multicolumn{1}{r}{\revision{18}}	  \\
	
	\midrule
	$t_{\text{pre}}$ & \multicolumn{1}{r}{1.73 }  & \multicolumn{1}{r}{1.88 }  & \multicolumn{1}{r}{3.74 } & \multicolumn{1}{r}{4.31}  & \multicolumn{1}{r}{ 3.06 } & \multicolumn{1}{r}{\revision{4.61}} \\

	$t_{\text{opt}}$ & \multicolumn{1}{r}{\revision{3.68}} & \multicolumn{1}{r}{\revision{6.01}}  & \multicolumn{1}{r}{\revision{6.13}}&	\multicolumn{1}{r}{\revision{5.71}} & \multicolumn{1}{r}{\revision{4.41}}& \multicolumn{1}{r}{\revision{5.86}} \\
	
	$t_{\text{pla}}$ & \multicolumn{1}{r}{\revision{0.21}} & \multicolumn{1}{r}{\revision{0.51}}& \multicolumn{1}{r}{\revision{0.96}}&	\multicolumn{1}{r}{\revision{0.87}}  & \multicolumn{1}{r}{\revision{0.27}}  & \multicolumn{1}{r}{\revision{0.47}}  \\
	
	\midrule
	\revision{$\varepsilon_{\text{geo}}$}	 &	\multicolumn{1}{r}{\revision{8.37}}&	\multicolumn{1}{r}{\revision{10.61}} &	\multicolumn{1}{r}{\revision{6.97}}& 	 \multicolumn{1}{r}{\revision{7.04}}  & 	\multicolumn{1}{r}{\revision{8.54}} & 	\multicolumn{1}{r}{\revision{6.23}}\\ 
	
	\midrule
	$\text{RMS}~\Delta$   &	\multicolumn{1}{r}{\revision{0.25}}     &	\multicolumn{1}{r}{\revision{0.36}}  &	\multicolumn{1}{r}{\revision{0.48}}&  \multicolumn{1}{r}{\revision{0.29}}  & 	\multicolumn{1}{r}{\revision{0.38}}  & 	\multicolumn{1}{r}{\revision{1.45}}  \\
	
	$\text{max} ~\Delta$	 &	\multicolumn{1}{r}{\revision{0.57}}&	\multicolumn{1}{r}{\revision{0.64}} &	\multicolumn{1}{r}{\revision{1.79}}& 	 \multicolumn{1}{r}{\revision{0.52}}  & 	\multicolumn{1}{r}{\revision{0.68}}  &\multicolumn{1}{r}{\revision{7.24}}  \\
		
	$E_{\msub{shape}}$	 &	\multicolumn{1}{r}{\revision{1.00}}&	\multicolumn{1}{r}{\revision{1.00}} &	\multicolumn{1}{r}{\revision{0.88}}& 	 \multicolumn{1}{r}{\revision{0.94}}  & 	\multicolumn{1}{r}{\revision{1.00}}  & 	\multicolumn{1}{r}{\revision{0.45}}  \\
						
\end{tabular}
\end{table}

Convergence times of our grid algorithm $t_{\text{opt}}$ mainly depend on the number of grid members and the mesh resolution. Precomputation times $t_{\text{pre}}$ depend on the mesh resolution, they include computing the distance fields and checking the boundary. Finally, planarization times $t_{\text{pla}}$ depend on the number of values of $\mathbf{t}=[t_1, ... t_k]$.

\subsection{Implementation}
Our grid design algorithm is implemented in \matlab, we use its GA-solver to solve the Optimization Problem~\eqref{eq:optimization}. To planarize the grid, we solve the Optimization Problem~\eqref{eq:planarization} utilizing \matlab's sequential quadratic programming solver using analytical gradients. We furthermore implemented the DER-simulation in C\texttt{++}, building upon the framework of \cite{Vekhter2019}. To compute the distance fields on the mesh, we use the VTP algorithm by \cite{qin16}. For the final computation of the geodesic paths, we use the algorithm for exact geodesics between two points by \cite{surazhsky05}.

\section{Discussion and Conclusions}\label{sec:discussion}

\subsection{Discussion and Limitations}
\paragraph{Surfaces with Inner Bumps}
We have shown that surfaces with inner bumps ($H<0$) can be realized without inner supports (cf. Figure \ref{fig:results}, example Hills), but in general, this is not possible (cf. Figure \ref{fig:results}, example Flower). 
\revision{In such cases, our shape stability energy  $E_{\msub{shape}} \in[0,1]$ provides information whether the grid can support itself or is likely to deviate from the target surface.
}

\paragraph{Holes in the Design Surface}
The input surfaces for our method need to be topological disks; we cannot compute grids on surfaces with holes. To adapt our approach to a surface with only one hole would require a second dual space because two boundaries must be taken into account. Additionally, a connection between the dual spaces needs to be established for geodesics that start at one boundary and end at the other. However, we can realize such surfaces using our present approach by splitting them into two or more non-convex surfaces~\cite{pillwein2021}.

\revision{
\paragraph{Deployment}
The deployment of physical grids depends on one rotational degree of freedom (DoF), which controls the expansion of the scissor linkage, and additional translational DoF, introduced by the notches. Kinematically, the grids are related to other elastic scissor-like grid structures which use notches \cite{Pillwein2020,Pillwein2020a,pillwein2021}, in contrast to structures with a rotational DoF only \cite{Panetta2019, Soriano2019}.

Our approach does not explicitly control individual notch lengths, it minimizes the overall notch lengths. Hence, when expanding a grid, it may not simply buckle into the final shape but nevertheless deliver perfect results when fixed to the supports. It would be interesting to adapt the planarization w.r.t.  producing a large number of notches of length zero and see how the shape of the planar grid changes.
}

\paragraph{Grid Optimization and Combinatorics}
In our current approach, we only change the combinatorics of the grid in the case of non-convex boundaries. 
Additional changes in the combinatorics during the optimization process, regardless of convexity, could promise even better results w.r.t. the grid energy $E_{\msub{grid}}$. However, we do not expect substantial improvements. This problem could be tackled using mixed-integer programming in the grid optimization, using an additional integer variable controlling the combinatorics.

\paragraph{Computational Efficiency}
\revision{Our method is streamlined for computational efficiency and delivers results in a matter of seconds.} We compute distance fields at the beginning and subsequently reuse them for all computations, even to circumvent costly path tracing of geodesics. 
\revision{Tracing geodesics using the implementation of Surazhsky \etal \shortcite{surazhsky05} significantly slows down the convergence of the grid optimization ranging from a factor of 108 for the example Starship up to 283 for the example Hills.
Despite our simplifications and the reduced resolution of the grid members, the effort energy $E_{\msub{effort}}$ is smooth and jumps only if grid combinatorics change (cf. Figure \ref{fig:effort_plots}). The shape stability energy $E_{\msub{shape}}$ is less smooth but still provides reliable information on whether the grid will be able to maintain the desired shape.}

\paragraph{Mesh Resolution} 
Our method requires a certain minimum resolution of the input mesh. If it is too coarse, we cannot find positions of the intersections of geodesics accurately (cf. Section \ref{sec:members_resolution}). Moreover, as the used distance fields only store distances for mesh vertices, we need interpolation, which depends on the resolution.
We noticed that the regions close to the boundary are prone to inaccuracies, so we recommend high resolution in this area.
For our examples, we used meshes with about 4000 vertices (cf. Table \ref{tab:commands}).

\subsection{Conclusions}\label{sec:conclusions}
We presented an approach for the computational design of elastic gridshell structures that
ensures to capture the surface characteristics well and provide aesthetic grid layouts.
Our fast form-finding algorithm is based on the notion of \textit{least-effort} and \textit{most-effort} geodesics, \revision{ self-supporting members, and mutual stabilization to} ensure close approximation of the input surfaces. \revision{An important outcome of our research is that even undulating surfaces can be well approximated by curve networks which minimize and maximize their normal curvature along their trajectories simultaneously.} In the future, we want to analyze this dependency further in order to provide theoretical results. 

Our method is based on distance computations only and omits expensive computations, like geodesic path tracing or physical simulations. 
We allow for surfaces with non-convex boundaries and updates of the combinatorics of the grid during optimization to ensure functional grids, which can be solved efficiently in the discrete domain. 
We introduce a planarization algorithm for geodesic grids with a quadratic objective function and quadratic constraints for the sake of efficient computation. As a result, the grids have straight members and are perfectly planar, which favors fabrication, transportation, and assembly.

Our method is inspired by architecture and design and intended as an easy-to-handle tool for designers to quickly create physically sound and aesthetically pleasing spatial grid structures.
Finally, we introduced a digital fabrication pipeline and presented a set of examples with varying curvatures, and produced small-scale gridshells as proof of our concept.

\begin{acks}
This research was funded by the 
\grantsponsor{1}{Vienna Science and Technology Fund (WWTF)}{https://www.wwtf.at}
\grantnum[https://www.wwtf.at]{1}{ICT15-082}.  
The authors thank Florian Rist and Johanna K\"ubert for fabricating the models, as well as Kurt Leimer and Dominik Pint for support with coding and production tasks.
\end{acks}

\bibliographystyle{ACM-Reference-Format}
\bibliography{elasticgrids2}


\end{document}